\documentclass[letterpaper]{jpconf}
\pdfoutput=1 
\usepackage{graphicx}

\newcommand{\bbbar}{\mbox{$\mathrm{b\overline{b}}$}}

\newcommand{\um}{\ensuremath{\rm \mu m}}

\begin{document}
\title{LHCb commissioning and readiness for first data}

\author{Helge Voss}

\address{Max-Plank-Institute f\"ur Kernphysik, Heidelberg, Germany}

\begin{abstract}
LHCb has been installed by spring 2008, followed by intensive
testing and commissioning of the system in order to be ready for first
data taking. Despite the horizontal geometry of the LHCb detector it 
was possible to collect over one million useful cosmic events that allowed a
first time alignment of the sub-detectors. Moreover events from
beam dumps during the LHC synchronisation tests provided very useful
data for further time and spacial alignment of the detector. Here we
present an overview of our commissioning activities,
the current status and an outlook on the startup in 2009.
\end{abstract}

\section{Introduction}
\label{intro}
The LHCb detector~\cite{DetectorPaper} at the Large Hadron Collider is
dedicated to study the physics in the decay of b-flavoured and other
heavy hadrons. At the nominal luminosity of $2 \times 10^{32}
\rm cm^{-2} s^{-1}$ at the location of LHCb and a production cross
section of $\approx 500\mu b$ at 14\,TeV proton-proton collisions,
$10^{12}$\bbbar\ pairs are expected to be produced annually. The
modest luminosity requirement for LHCb can be met very early during 
LHC operation, long before the high luminosity runs and later maintained
adjusting the  LHC optics when larger luminosities are delivered to the other
multi-purpose experiments at the LHC.

\section{Detector Overview}
\begin{figure}[ht!]
\begin{center}
\includegraphics[width=0.55\textwidth]{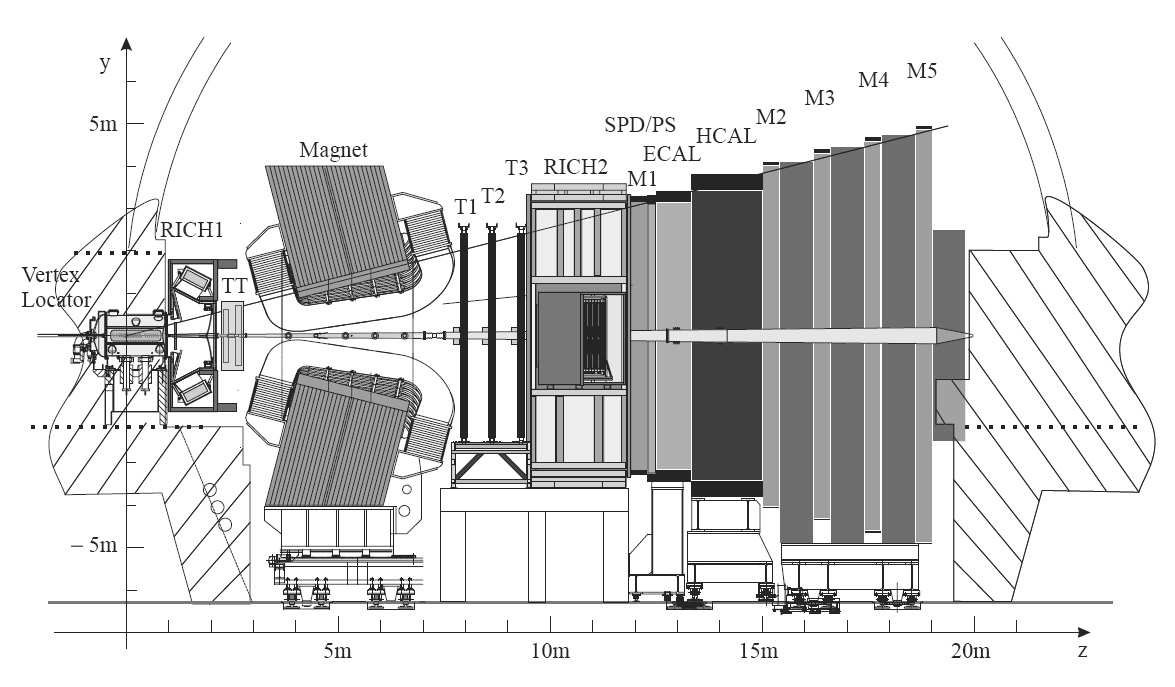}
\includegraphics[width=0.3\textwidth]{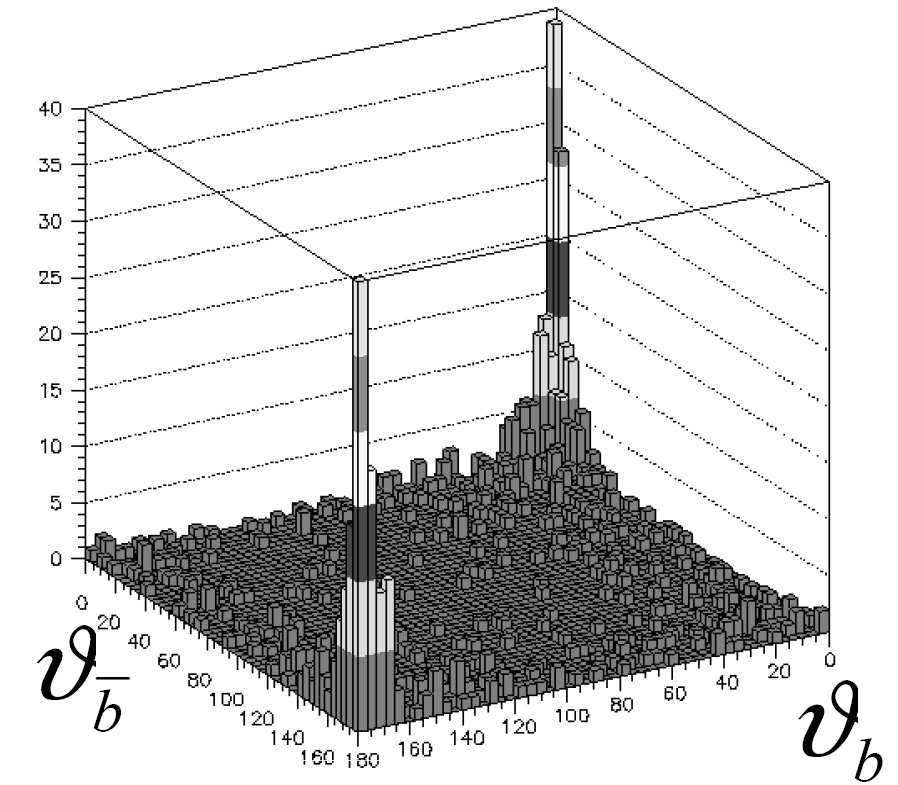}
\end{center}
\caption{Left: Overview over the LHCb detector. Right: the expected
  angular distribution of \bbbar\ pairs produced at LHC.}
\label{fig:lhcb}
\end{figure}

LHCb is designed as a single arm forward spectrometer
adapted to the angular distribution of the \bbbar\ pairs which are
produced predominantly at low polar angles as shown in
Figure~\ref{fig:lhcb}.
The detector covers an angular region from about 10\,mrad to
300\,(250)\,mrad in the bending (non-bending) plane of the magnet.
LHCb uses a warm dipole magnet that delivers an integrated magnetic
field of 4\,Tm. There are two Ring Imaging Cherenkov detectors, RICH1
and RICH2 located before and behind the magnet, respectively. These
two detectors that use three different radiator materials, aerogel,
$\rm C_4F_{10}$ and $\rm CF_4$, give excellent $\pi$-K separation in a
large momentum range of 2-100 GeV/c. The vertex detector (VELO) and
the full silicon tracking station (TT-station) before the magnet and
the three tracking stations behind the magnet provide the tracking
system of LHCb.  The stations behind the magnet are divided into the
silicon strip Inner Tracker and the straw tube Outer Tracker. The
calorimeter system consists of a Scintillator Pad Detector and
Preshower (SPD/PS), a ``shashlik type'' electromagnetic calorimeter,
and the Fe plus scintillating tile hadronic calorimeter. The detector
is completed with a muon system that uses multi-wire proportional
chambers and triple-GEM's in the very first muon station.

LHCb is using a level-0 hardware trigger that reduces the initial interaction
rate of 16\,MHz down to 1\,MHz using transverse energy measured
in the calorimeters and the two highest $P_T$ muons seen in the muon chambers.
After a level-0 trigger, the whole detector is read out and the information
used in the high level trigger (HLT) where in several stages of reconstructing
the events, the rate is reduced to 2\,kHz at which the events are stored.

\section{Commissioning}
Commissioning activities have started in 2007 with the first individual 
sub systems and safety devices being tested in parallel with
the final installation of subdetectors which finished in spring 2008. First
commissioning activity made use of internal test pulses to test the
functionality of the readout and of detector elements. Sending test
pulses and receiving the detector responses allows testing the full
control and readout chain of the sub-detectors.  Dead or noisy channels
are located (and eventually fixed/replaced where possible) and
consistency checks also allow in certain cases to track down possible
cabling swapping and alike. From known cable lengths and delays
measured using test pulses the initial settings for the time alignment
were set for the different sub-detectors. 

\subsection{Commissioning With Cosmic Events}
\begin{figure}[ht!]
\begin{center}
\includegraphics[width=0.45\textwidth]{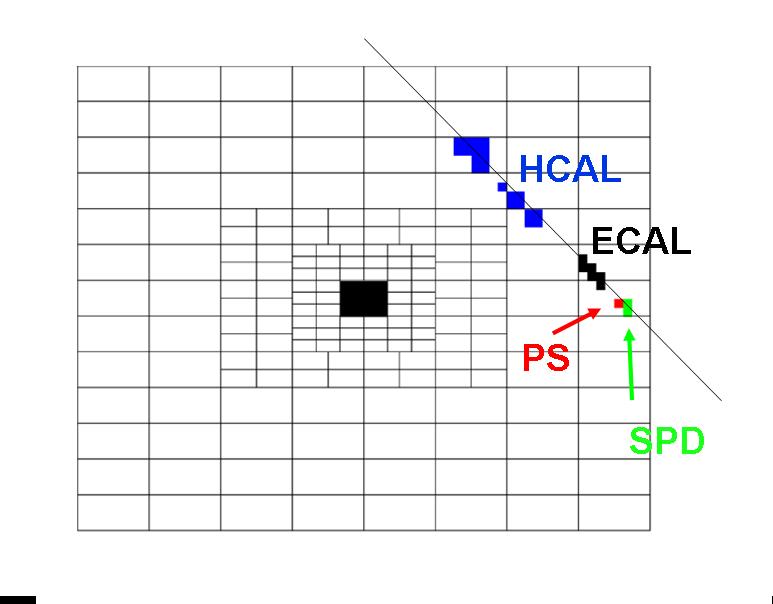}
\includegraphics[width=0.5\textwidth]{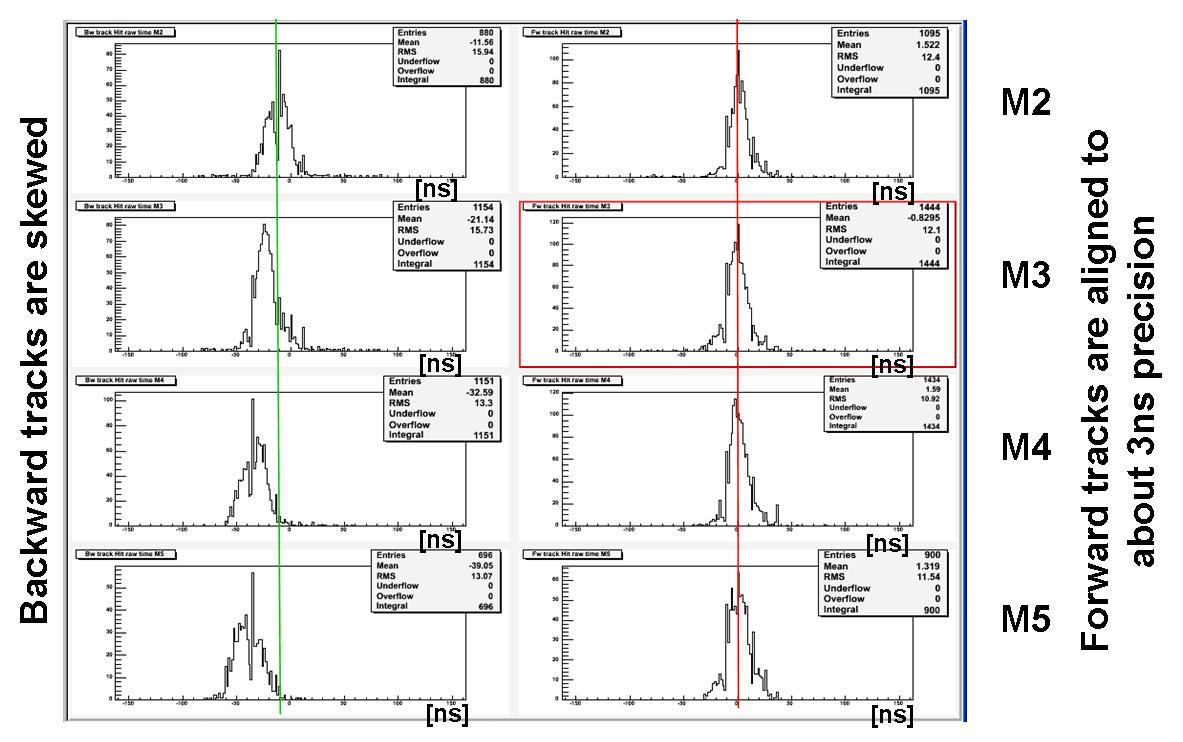}
\end{center}
\caption{Left: a cosmic event in LHCb as seen by the calorimeters.
  Right: time alignment of the muon chambers with cosmic particles.
  One can see that a time alignment with a precision of about 3\,ns is
  achieved for particles that cross the detector in the proper
  direction, while for backward going particles the signal arriving
  times are skewed.}
\label{fig:cosmic}
\end{figure}

Despite the horizontal geometry of the LHCb detector and a very low
rate under 1\,Hz of reasonably horizontal cosmic particles within the
250\,mrad of the horizontal acceptance, over a million cosmic events
were recorded in LHCb. First these rare cosmic events were used to
commission the level-0 trigger, aligning calorimeters and later the
muon chambers in time to one another as done end of 2007. Special low
thresholds were used in the calorimeter to be able to trigger on MIPs
together with a loose coincidence between ECAL and HCAL which in the
end results in a trigger rate of about 10\,Hz. A cosmic particle showing a nice
``track'' in the calorimeters is shown in Figure~\ref{fig:cosmic}.
Internal time alignment of the various calorimeter cells using the
known pulse-shape from test-beam measurements is reaching a precision
of about 3\,ns.
The muon trigger also was used for cosmic events using a
coincidence between two stations omitting the vertex constrained used
in the real experiment's trigger. The muon stations were also time
aligned using these cosmic events as shown in the left part of Figure~\ref{fig:cosmic}.
Time alignment with cosmic muons in LHCb needs to disentangle the
forward from the backward going particles. Which is nicely 
displayed in Figure~\ref{fig:cosmic}, showing the timeing for
forward and backward particles.

The Outer Tracker and Inner Tracker stations located in front of the
calorimeter stations (separated only by the RICH2 detector) were 
commissioned using the cosmic particles triggered by either the
calorimeters or the muon detector. Despite the small surface of the
Inner Tracker, some clear cosmic events could be seen as
``coincidences'' of hits in neighbouring silicon layers within a
detector box. This allowed for a very first rough time alignment of
the Inner Tracker with the calorimeters, using known cable lengths for
the internal time alignment. Very helpful for this exercise was the
possibility in LHCb to trigger on and read out up to 15 consecutive
events. This allowed to read out a window of $\pm$7 events
around an event triggered by the calorimeters and find the
corresponding coincidences in the Inner Tracker. Two tracks with hits
in all three Inner Tracker stations were found in over a million
cosmic triggers. 

\subsection{Commissioning With Beam Events}
\begin{figure}[ht!]
\begin{center}
\includegraphics[width=0.28\textwidth]{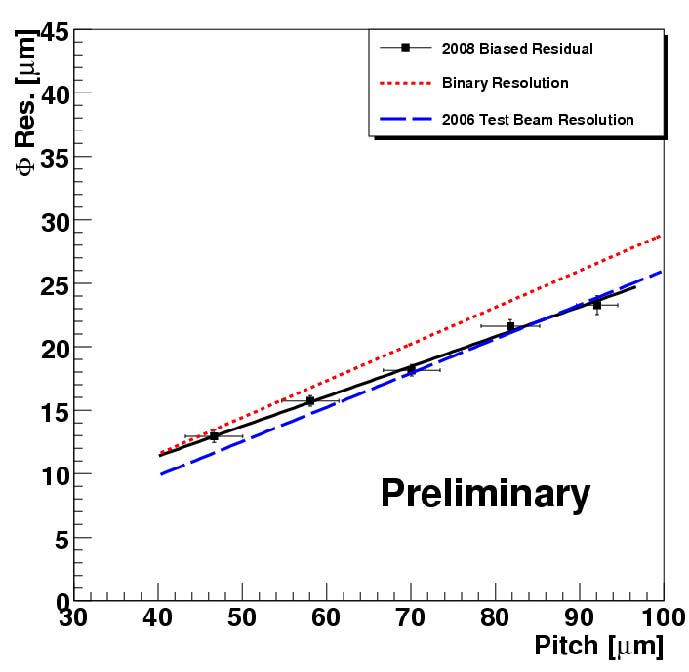}
\includegraphics[width=0.38\textwidth]{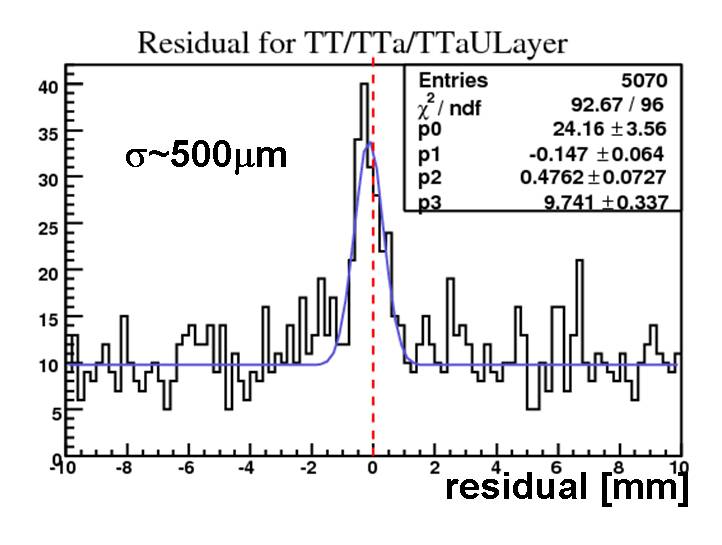}
\end{center}
\caption{Spatial alignment using beam dump events. The right plot shows the 
  resolution in $\Phi$ of the VELO as a function of the strip pitch
  which changes with sensor position.  The left plot shows the
  residual of tracks reconstructed in the VELO and extrapolated to one
  of the TT-station layers.}
\label{fig:VeloTT}
\end{figure}

While commissioning with cosmic events is not feasible for the vertex
detector and the TT-station ``far away'' from the calorimeters, we
profited here from synchronisation tests in the injection lines done
by the LHC machine group. During these tests, the injection beam of
the SPS accelerator was dumped on a beam stopper (TED) located about
300m downstream of LHCb. This resulted in a large number of about
10\,GeV muons in LHCb and the first tracks seen in the VELO. Careful
analysis of these data allowed also for first spatial alignment
studies of the VELO, and some of those results are shown in
Figure~\ref{fig:VeloTT}. It shows the hit residuals as a function of
the strip pitch which increases with radius of the radial strips. The
plot shows the performance achieved during the test-beam measurements
and the expected performance for binary resolution without any charge
sharing between the strips. The current measured resolution after this
first alignment step was already well inbetween these two lines.

The TT-station consists of only four individual layers, too little to
make stand alone tracking. However, once can extrapolate the VELO
tracks and look at the hit residuals in the different layers of the
TT-station.  This results in very nice peaks with a widths of about
500\,\um\ and small overall shifts of the order of some hundred \um\
depending on the layer.  For one layer this is shown in
figure~\ref{fig:VeloTT}.  The spread expected from simple extrapolation
of the tracks would be of the order of 300\,\um due to the resolution
in the VELO.  This shows a very good initial alignment of the two
sub-detectors w.r.t each other as well as a first hint of the internal
positioning accuracy of the TT-station modules.

\begin{figure}[bt!]
\begin{center}
\includegraphics[width=0.35\textwidth]{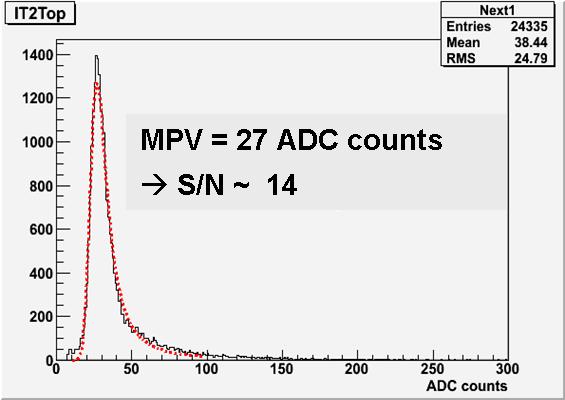}
\includegraphics[width=0.35\textwidth]{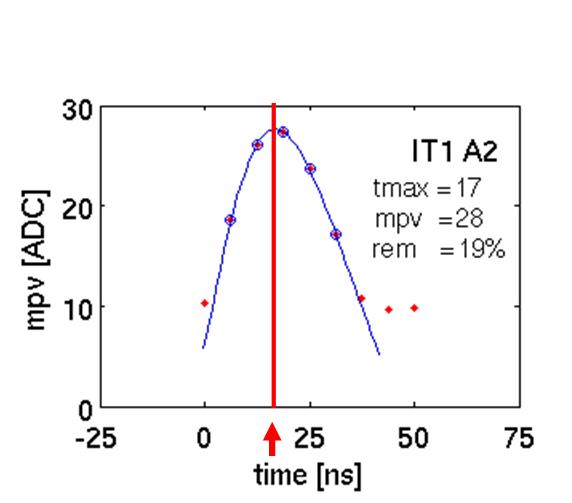}
\end{center}
\caption{Time alignment of the Silicon Trackers using beam dump events. The left plot
shows a Landau distribution from the deposited charge in ADC counts and the right
plot shows the reconstructed pulseshape.}
\label{fig:pulseshape}
\end{figure}
The large number of hits recorded in the silicon trackers during these ``TED-events'' 
also allowed for a very accurate time alignment of the detectors within a few ns, not
achievable with the very limited number of cosmic events. A reconstructed charge distribution 
in the Inner Tracker and a pulseshape derived from varying the relative timing between the
beam and the readout clock is shown in Figure~\ref{fig:pulseshape}.

\section{Conclusions}
Commissioning in LHCb had started as early as 2007.  making extensive
use of internal generated test pulses but also cosmic ray events
despite the horizontal geometry of LHCb. Very useful proved the beam
induced events when the injection beam was dumped in the injection
line some 300\,m behind the detector which in the end allowed us to get
first time alignment and spatial alignment of basically all
sub-detectors. The long shutdown that followed after the LHC incident
on September 19 is being used for maintenance and further improvement on the
overall efficiency of the detector. The time is also being used for
more tests of the detector readout at the full trigger rate of 1\,MHz,
data processing and writing. The latter has been successfully tested
to speeds up to 1.9\,kHz, close to the design of 2\,kHz. For the year
2009 run, the full HLT computer farm will be available which will
allow to take data at full LHCb rate.

\end{document}